\documentclass[journal,twoside]{IEEEtran}
\usepackage{amsmath,amsfonts}
\usepackage{algorithmic}
\usepackage{algorithm}
\usepackage{array}
\usepackage[caption=false,font=normalsize,labelfont=sf,textfont=sf]{subfig}
\usepackage{textcomp}
\usepackage{stfloats}
\usepackage{url}
\usepackage{verbatim}
\usepackage{amssymb}
\usepackage{mathtools}
\usepackage{graphicx}
\usepackage{cite}
\usepackage{hyperref}
\usepackage{xcolor}
\usepackage{multirow}

\begin{document}
\bstctlcite{IEEEexample:BSTcontrol}
\title{Towards robust quantitative photoacoustic tomography via learned iterative methods}

\author{Anssi Manninen, Janek Gröhl, Felix Lucka, and Andreas Hauptmann~\IEEEmembership{Senior Member,~IEEE}
\thanks{AM and AH were supported in part by the Research council of
Finland (Project No. 353093, Finnish Centre of Excellence
in Inverse Modeling and Imaging; and Project No. 359186, Flagship
of Advanced Mathematics for Sensing Imaging and Modelling; Projects No.
338408, 359915, Academy Research Fellow)}
\thanks{Codes published at: \url{https://github.com/AnsManni/Learned_iterative_QPAT}}
\thanks{A. Manninen is with the Research Unit of Mathematical Sciences,
University of Oulu, Oulu, Finland.}
\thanks{J. Gröhl is with ENI-G, a Joint Initiative of the University Medical Center Göttingen and the Max Planck Institute for Multidisciplinary Sciences, Göttingen, Germany}
\thanks{F. Lucka is with the Centrum Wiskunde \& Informatica, Amsterdam, The Netherlands.}
\thanks{A. Hauptmann is with the Research Unit of Mathematical Sciences,
University of Oulu, Oulu, Finland and with the Department of Computer
Science, University College London, London, United Kingdom.
}
}

\maketitle

\begin{abstract}
Photoacoustic tomography (PAT) is a medical imaging modality that can provide high-resolution tissue images based on the optical absorption. Classical reconstruction methods for quantifying the absorption coefficients rely on sufficient prior information to overcome noisy and imperfect measurements. As these methods utilize computationally expensive forward models, the computation becomes slow, limiting their potential for time-critical applications. As an alternative approach, deep learning-based reconstruction methods have been established for faster and more accurate reconstructions. However, most of these methods rely on having a large amount of training data, which is not the case in practice. In this work, we adopt the model-based learned iterative approach for the use in Quantitative PAT (QPAT), in which additional information from the model is iteratively provided to the updating networks, allowing better generalizability with scarce training data. We compare the performance of different learned updates based on gradient descent, Gauss-Newton, and Quasi-Newton methods. The learning tasks are formulated as greedy, requiring iterate-wise optimality, as well as end-to-end, where all networks are trained jointly. The implemented methods are tested with ideal simulated data as well as against a digital twin dataset that emulates scarce training data and high modeling error.
\end{abstract}

\begin{IEEEkeywords}
Quantitative photoacoustic tomography, nonlinear inverse problems, deep learning, learned iterative methods, convolutional neural networks, digital twin.
\end{IEEEkeywords}

\section{Introduction}
 During recent decades, quantitative photoacoustic tomography (QPAT) has received increasing interest as a new, non-ionizing, \emph{in vivo} imaging modality due to its ability to produce quantitative images of optical parameters with higher resolution than purely optical modalities by combining the propagation of non-scattering ultrasound and optical contrast~\cite{ beard,wang2017photoacoustic,li2009photoacoustic}. In a typical QPAT setup, short pulses of light are emitted to a target region, where the light scatters and is absorbed, causing the emergence of ultrasound waves, which are measured at the boundary. The measurements carry information on chromophores that absorbed the light and can be used to form an inverse problem to reconstruct quantitative chromophore concentrations~\cite{cox2012quantitative,li2021recent}, revealing functional tissue properties such as blood oxygenation.

The inverse problem of QPAT is usually formulated as two separate problems: the acoustical and optical inversion. In the acoustic inverse problem, the objective is to recover the initial pressure field, which primarily provides qualitative structural information that can already carry diagnostic meaning in clinical applications, see, e.g.~\cite{xia2013small,wang2016practical}. Various established methods exist to solve it~\cite{beard,hauptmann2024model}, and we assume it to be solved in this work. While the acoustic waves propagate nearly scattering-free, scattering of photons cannot be neglected when solving the optical inverse problem. Therefore, the measured data is modeled to be nonlinearly dependent on the unknown optical parameters. For biological tissues, the commonly accepted model for light transport is the radiative transfer equation (RTE)~\cite{arridge1999optical}, which often has to be approximated for computational efficiency. The most popular of these is the diffusion approximation (DA)~\cite{cox2011multiple,tarvainen2013image}, which is most accurate in highly diffusive regions~\cite{tarvainen2013image,cox2011multiple,firbank1996investigation}. Alternatively, a popular choice is to stochastically estimate the photon transport with Monte Carlo methods~\cite{leino2019valomc}. 

The solution of the optical nonlinear inverse problem with a light propagation model can be formulated as an optimization problem. This nonlinear optimization problem is popularly solved by using information from second-order derivatives iteratively~\cite{nocedal1999numerical}. However, computing or storing the Hessian is often impractical for a large number of unknowns and hence it is often approximated with quasi-Newton or Newton type methods~\cite{nocedal1999numerical,saratoon2013gradient}.

The reconstruction of the weakly data-dependent and highly ill-posed scattering is the biggest challenge. To overcome it, one could assume constant scattering values and only estimate the mildly ill-posed absorption~\cite{cox2005quantitative}. This can lead to poor absorption estimates if the assumed scattering is inaccurate. But even simultaneous estimation of absorption and scattering from a single illumination is non-unique. To guarantee uniqueness, multiple sources with different illumination patterns or multiple wavelengths need to be utilized~\cite{cox2011multiple,cox2009estimating}. 

\subsection{Deep learning-based techniques for QPAT} 
\begin{figure}[h!tbp]
\centering
\vspace{-0.45cm}
\includegraphics[width=0.23\textwidth]{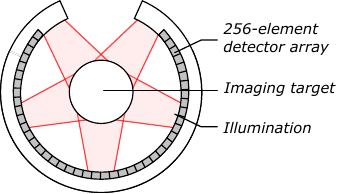}
\includegraphics[width=0.25\textwidth]{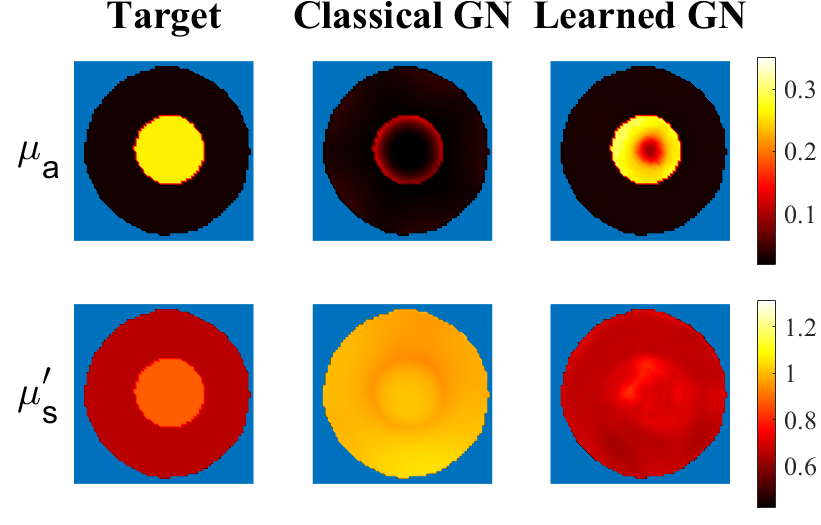}
\caption{Left: Setup of the photoacoustic hardware. Right: Absorption ($\mu_a$) and reduced scattering ($\mu_s^\prime$) reconstructions of a digital twin sample (left) after the optical inversion. Due to the huge modeling error, the classical Gauss-Newton solver (mid) is not able to reconstruct the targets robustly, whereas its deep-learned assisted variant (right) is.}
\label{fig:hook}
\end{figure}

 Solving the nonlinear optical problem with classical model-based techniques comes with difficulties~\cite{hauptmann2024model,tarvainen2024quantitative} that we illustrate in Fig.~\ref{fig:hook}). Recently, data-driven methods have been studied to overcome these. However, in contrast to the acoustic problem, where a wide range of methods have been studied~\cite{hauptmann2020deep,grohl2021deep}, approaches for the optical problem have been more limited due to the computational burden of including the imaging model.

Deep learning techniques developed for the optical problem of QPAT have focused primarily on learning a direct reconstruction, e.g., to fully learn a single-step reconstruction operator that takes multiple initial pressure images as input and directly and computationally efficiently estimates absorption coefficients~\cite{chen2020deep}, chromophore concentrations~\cite{cai2018end,yang2019quantitative} or blood oxygen saturation~\cite{bench2020toward,grohl2024distribution}. A downside of fully learned operators is that they do not incorporate an explicit imaging model, and the learning task is highly complex. They thus require a large quantity of training samples, and generalization is limited~\cite{grohl2023moving}.
For clinical applications, limited sample acquisition and missing ground truth for training remain a problem. Consequently, with only a handful of experimental samples the fully learned reconstruction methods are not likely to generalize well.

The idea of model-based learned methods is to incorporate an imaging model to reduce the complexity of the learning task. However, in contrast to the fully learned approach, modeling inaccuracies can now affect the result. Some recent model-based work has concentrated on accelerating classical iterative solvers by replacing the hand-crafted light transport model by a learned Fourier neural operator~\cite{liang2024deep,liang2025self}.
In the popular model-based learned iterative reconstruction methods, the classically computed model information and the neural networks updating the unknown values are iteratively repeated $K$ times, analogous to classical iterative optimization solvers~\cite{arridge2019solving}.
While the repeated model evaluation leads to longer run times, they have been shown to outperform single-step reconstructions in quality and generalizability with smaller network sizes~\cite{adler2018learned,hauptmann2018model,hammernik2018learning}. However, their application to nonlinear inverse problems is severely limited due to the additional model complexity. Specifically, the approach has not been used for QPAT and is limited to related modalities such as diffuse optical tomography~\cite{mozumder2021model} and electrical impedance tomography~\cite{herzberg2021graph,colibazzi2023deep,alberti2025deep}, often using a cheaper \emph{greedy} training regime of training each updating network (iterate) separately instead of the preferred \emph{end-to-end} training of all iterates usually employed in linear inverse problems~\cite{arridge2019solving}.

\subsection{Contributions of this work}
 The scientific literature on model-based iterative learned methods for nonlinear inverse problems remains scarce and 
several questions remain:
\begin{enumerate}
    \item Influence of the step directions, i.e., of the underlying iterative optimization method.
    \item Influence of the number of the learned iteratively updating networks on the reconstructions
    \item Performance under large modeling errors
    \item Training regimes: Greedy vs. end-to-end
\end{enumerate}
We provide insights into the above by investigating the performance of the model-based learned iterative method by prototyping it for the QPAT problem. QPAT is ideal for our purposes due to its nonlinear nature, and because there is a general need for fast and robust methods. Importantly, the optical problem involves solving the mildly ill-posed absorption and severely ill-posed scattering, demonstrating the performance on both types of unknowns. As the model-based iterative method requires repeated use of the light propagation model, the DA is used for training the networks, providing a possibility to investigate the model error compensation capabilities.

To gain insight into how reconstruction accuracy and computational times differ between network realizations, we test gradient descent, Gauss-Newton, and quasi-Newton rank-1-based learned updates with an increasing number of learned updating blocks. An additional focus is to compare performances between greedily and end-to-end trained networks. While all previous work on nonlinear inverse problems has utilized computationally cheaper greedy training, we also implemented end-to-end training for the first time and will examine if there is a significant performance difference between the two schemes. To compare each implemented method, we first consider a simulated 2D study with a large number of samples and low modeling error. To investigate the performances closer to experimental conditions, we utilize a 2D digital twin~\cite{grohl2023moving,grohl2025digital} with a scarce amount of training data, and a high modeling error when using the DA.

\section{Quantitative photoacoustic tomography}\label{sec:QPAT}
 In QPAT, the acoustic inversion is followed by the optical inversion, for which the data, absorbed energy density $h(r)$ for region $r\in\Omega \subset \mathbb{R}^n$ $(n=2$ or 3), is acquired from the acoustical problem via the relation $h(r)=\frac{p_0(r)}{\hat{\Gamma}(r)},$ where $p_0$ is the initial pressure and $\hat{\Gamma}$ is Gr\"uneisen parameter. In this paper, only the optical inverse problem is covered. Therefore, we assume that the initial pressure field is already solved and divided by the known Gr\"uneisen parameter, yielding an estimate of the absorbed optical energy density $h$.

\subsection{Forward model of the optical problem}\label{sec:forward}
 For the optical inversion, we consider a discretization of the domain with $P$ non-overlapping finite elements and $N$ grid coordinates. The optical parameters in $\Omega$ are then represented in the basis
\begin{equation}\label{eq:basis}
    \begin{aligned}
\mu_a(r) \approx \sum_{t=1}^N \mu_{a_t} \varphi_t(r), \quad
\mu_s(r) \approx \sum_{t=1}^N \mu_{s_t} \varphi_t(r),
\end{aligned}
\end{equation}
where $\mu_a(r)$ is the absorption field, $\mu_s(r)$ is the scattering field and $\varphi_k(r)$ is the finite element basis function. The absorbed optical energy density is linked to the scattering and absorption coefficients via the relation
\begin{equation}\label{eq:E_density}
    h(r)=\mu_a(r) \Phi(\mu_a(r),\mu_s(r)),
\end{equation}
where $\Phi(r)$ is the fluence in $r \in \Omega$. The fluence depends on the absorption $\mu_a(r)$ and scattering $\mu_s(r)$ distributions in $\Omega$ in a nonlinear way. For the fluence, the same basis is used as for the optical coefficients in \eqref{eq:basis}. To solve the dependency between the optical parameters and the fluence in \eqref{eq:E_density}, a light propagation model is introduced. The commonly accepted model for biological medium is the RTE, originating from Maxwell's equations~\cite{ishimaru1978wave}. For QPAT, the time-independent RTE can be found in the $A$ section of the supplementary file.

Due to the high computational cost of solving the RTE, a diffusion approximation (DA) has been popularly applied in biomedical applications. As in this work (see Sec. \ref{sec:learned_QPAT}), we will numerically test the model-based learned iterative methods, which evaluate the light propagation model multiple times in each training step. Therefore, it is practical to consider the DA for light propagation. However, the DA can only predict the photon fluence accurately if the domain is approximately diffusive, that is, the scattering needs to be significantly larger than the absorption, and the domain size should be sufficiently large compared to the average scattering length. The exact form of the diffusion approximation and the used boundary condition can be found in Section B of the supplementary file. For more information on the diffusion approximation and its derivation, see, e.g.~\cite{arridge1999optical}.

To numerically solve the DA, the finite element (FE) approximation (for details, see~\cite{malone2015reconstruction,tarvainen2012fem}) is applied. Applying the FE approximation in the chosen basis $\varphi_t$ yields a matrix form for the DA from which the fluence can be computed as
\begin{equation}\label{eq:fluence}
    \Phi = (M+C+R)^{-1}Q.
\end{equation}
The exact forms of matrices $M, C, R$, and vector $Q$ can be found in the $B$ section of the supplementary file.


\subsection{Variational methods for solving the optical inversion}
 To define the inverse problem for $I$ illuminations, let us consider a discrete observation model of the form
\begin{equation}\label{eq:obs_model}
h = F(x)\:+\:e,
\end{equation}
where $h\in \mathbb{R}^{N\times I}$ is the absorbed optical energy density at locations $r_t$, $t=1,\dots, N$, $x\in\mathbb{R}^l$ contains the unknown scattering and absorption parameters, $F:\mathbb{R}^l\to\mathbb{R}^{N\times I}$ is a nonlinear forward operator mapping the optical parameters $x$ to the data space, and $e\in\mathbb{R}^{N\times I}$ is the additive measurement noise. In this work, the operator $F$ corresponds to the RHS of the equation \eqref{eq:E_density}, where the fluence $\Phi$ is obtained from the finite element solution of the DA in the form of \eqref{eq:fluence}. The inverse problem is then to recover the set of unknown optical parameters $x$ from the measured noisy data $h$. 

A well-established framework to solve the inverse problem of the nonlinear observation model \eqref{eq:obs_model} is the variational approach. In the variational approach for QPAT, the optical parameters $x$ are solved by minimizing the functional
\begin{equation}\label{eq:min_problem}
    \mathcal{E}(x,h)=\frac{1}{2}\|L_e(h-F(x))\|_2^2+\alpha \Psi(x),
\end{equation}
where $L_e$ is the weighting matrix, $\alpha$ is the regularization parameter, adjusting the importance of the regularizer $\Psi$, which encodes the prior information about $x$. By using the regularizer $\Psi(x)$, the solution space of $x$ can be restricted such that smooth or sparse solutions are promoted.
To guarantee the existence of a set $x$ that uniquely describes data $h$, it is necessary to either use multiple illuminations with different patterns or approximate the wavelength dependency of the scattering. The wavelength dependency in biological tissue has been approximated with an exponentially decaying function such as $f(\lambda)=\lambda^{-b}$, where $b \in \mathbb{R}$ is determined from experiments~\cite{jacques2013optical}.

A popular approach to minimize the functional \eqref{eq:min_problem} is to use an iterative optimization method of the form 
\begin{equation*}
    x^{(k+1)} = x^{(k)} + \beta_kp_k\left(x^{(k)},h\right),
\end{equation*}
where $p_k$ is the update direction and $\beta_k$ is the step length. The initial optical values $x^{(0)}$ need to be set according to prior knowledge from experiments. For a differentiable functional $\mathcal{E}$, the update direction $p_k$ can utilize information from the first-order derivatives 
\begin{equation*}
\nabla\mathcal{E}\left(x^{(k)},h\right)=J_k^TW_e(h-F(x^{(k)})) + \alpha \nabla\Psi(x^{(k)}), 
\end{equation*}
forming the gradient descent (GD) update 
\begin{equation}\label{eq:GD_update}
    x^{(k+1)} = x^{(k)} + \beta_k\nabla{\mathcal{E}}\left(x^{(k)},h\right).
\end{equation}
where $J_k$ is the Jacobian of the $F$ at $x^{(k)}$, $\nabla\Psi(x^{(k)})$ is gradient of the regularizer, and $W_e=L_e^TL_e$.

Alternatively, the second-order derivatives can be computed to utilize the Newton's update direction
\begin{equation}\label{eq:GN_update}
    x^{(k+1)} = x^{(k)} + \beta_kH_k^{-1}\nabla\mathcal{E}\left(x^{(k)},h\right),
\end{equation}
where $H_k$ is the Hessian of $\mathcal{E}$ at $x^{(k)}$.  
As it is rarely computationally feasible to solve the exact second derivatives, the Gauss-Newton (GN) update
\begin{equation}\label{eq:quasi_update}
    x^{(k+1)} = x^{(k)} + \beta_k(J_k^{\rm T}W_eJ_k + H_\Psi)^{-1}\nabla{\mathcal{E}\left(x^{(k)},h\right)},
\end{equation}
is used instead, where the Hessian of $F$ is now approximated with $J_k^{\rm T}WJ_k$ and $H_\Psi$ is the Hessian of the regularizer at $x^{(k)}$. Alternatively, the inverse of the Hessian $H_k^{-1}$ can be approximated with quasi-Newton methods~\cite{nocedal1999numerical}.
In these methods, the approximations of $H_k^{-1}$ try to encapsulate curvature information of the exact Hessian on $x^{(k)}$ based on the gradients.

As the used FE approximation of the DA and hence $F(x)$ is differentiable, the differentiability of the whole functional $\mathcal{E}$ depends on the type of regularizer $\Psi(x)$ used. A popular, differentiable regularizer is the weighted $\ell^2$-norm 
\begin{equation}\label{eq:el_two}
    \Psi(x) = \frac{1}{2} \|L_x(x-x_\mu)\|^2_2,
\end{equation}
with weighting matrix $L_x\in \mathbb{R}^{n\times n}$ and mean $x_\mu\in \mathbb{R}^l$. The gradient $\nabla\Psi(x^{(k)})$ is then simply $L_x^T L_x(x-x_\mu)$. 
The weighting matrix $L_x$ can be used to introduce spatial correlations between the nodes, for example, by using the Ornstein-Uhlenbeck process
\begin{equation}\label{eq:O-U}
    \left(\Gamma_{x} \right)_{p,t} = \sigma^2 \exp \left(-\frac{\left\|r_p-r_t\right\|}{\ell}\right),
\end{equation}
where $r_p$ and $r_t$ are the locations of the $p$th and $t$th nodes, $\sigma^2$ is the variance, and $\ell$ is the characteristic length scale controlling the spatial decay of the correlation. The weighting matrix $L_x$ is then the Cholesky factor of the inverse $\Gamma_x^{-1}$. An example of a non-smooth regularizers is given by the (discretized) total variation (TV), 
\begin{equation}\label{TV_seminorm}
   T V(x)=\sum_{e=1}^E d_k\left|\boldsymbol{x}_{p(e)}-\boldsymbol{x}_{t(e)}\right|,
\end{equation}
where $d_k$ is the length of the $k$th edge in the mesh used, between nodes $x_p$ and $x_t$, and $E$ is the total number of edges in the mesh. The gradients for the update rules \eqref{eq:GD_update}-\eqref{eq:quasi_update} are now not defined, and to minimize \eqref{eq:min_problem}, more complex iterative methods need to be used.

The convergence of the introduced classical iterative methods depends on the selection rules of the step length $\beta_k$, which can be challenging and computationally slow to find for nonlinear problems \cite{kaltenbacher2008iterative}. Also, note that, for a nonlinear minimization problem, such as \eqref{eq:min_problem}, there are no general methods to choose a suitable regularization parameter $\alpha$, and it is often determined by hand-tuning.

\subsection{Previous learned reconstructions for the optical inversion}
 Let us shortly summarize previous deep learning approaches to solve the quantitative problem. Majority if approaches have considered a fully learned approach, that means a neural network is trained to predict optical values, or often blood oxygenation ($\mathrm{sO}_2$), directly from either acoustic measurement data or an intermediate reconstruction, such as reconstructed initial pressure $p_0$. 
More precisely, let $[p_0]_\lambda$ be a set of a multi-wavelength reconstructions, and denote by $\mathrm{sO}_2$ a spatially resolved $\mathrm{sO}_2$-map. Then the a neural network $\Lambda_\theta$ with parameters $\theta$ is trained to estimate
$\Lambda_\theta([p_0]_\lambda) \approx \mathrm{sO}_2$,
see for instance~\cite{bench2020toward,grohl2021deep}. 

Alternatively, the network can be trained to estimate the optical values instead of blood oxygenation. Here, we will consider as baseline for our proposed approach a network that predicts the optical values $[\mu_a,\mu'_s]$ from the absorbed energy density, that is
\begin{equation}\label{eq:baselineUnet}
\Lambda_\theta(h)\approx [\mu_a,\mu'_s].
\end{equation}
While such a fully learned approach works well for in distribution data, it often generalizes insufficiently well for out-of-distribution data and especially experimental measurement data.

\section{Model-based learned iterative methods for the optical problem}\label{sec:learned_QPAT}
 Classical variational methods to compute solutions of \eqref{eq:min_problem} often heavily depend on the prior information encoded in the regularizer $\Psi$ and may need a large number of iterations to reach sufficient accuracy and hence limit time-critical applications. In the following, we will learn a reconstruction operator that encapsulates the prior information and learn effective update steps to improve reconstruction quality and speed.

For our application, in analogy to the classical updates \eqref{eq:GD_update}-\eqref{eq:quasi_update}, we can write the deep-learned iterations of the model-based learned iterative method for $K$ iterations as 
\begin{equation}\label{eq:learned_iterative_update}
    x^{(k+1)} = \Lambda_{\theta_k}\left(x^{(k)}, p_k\left(x^{(k)},h\right)\right),
\end{equation}
where  $k=0, \ldots, K-1$, and each iteration has an updating network $\Lambda_{\theta_k}$ with weights $\theta_k$. The reconstruction operator is then defined as the result after $K$ updates 
\begin{equation}\label{eq:recon_operator}
    \mathcal{R}_\theta(h):=x^{(K)},
\end{equation}
where
$\theta=\left\{\theta_0, \ldots, \theta_{K-1}\right\}$. The commonly used architecture for the network $\Lambda_{\theta_k}$ is based on the residual network ResNet~\cite{he2016deep}, which consists of several multi-channel convolutional layers with activation functions and the last single-channel layer giving the update to the previous reconstruction $x^{(k)}$. As the updating networks $\Lambda_{\theta_k}$ in \eqref{eq:learned_iterative_update} repeatedly receive information based on the model, the network sizes can now be kept relatively small. A detailed description of architecture used for $\Lambda_{\theta_k}$  is given in Section \ref{sec:archithecture}.   

\subsection{Choices for the update direction}\label{sec:step_choices}
 In our work, we will consider the model-based learned iterative solver \eqref{eq:learned_iterative_update}, where the directions $p_k$ are chosen as GD from \eqref{eq:GD_update}, i.e.,  
\begin{equation}\label{eq:learnedGDStep}
p_k\left(x^{(k)},h\right)=J_k^TW_e(h-F(x^{(k)}))
\end{equation}
as well as GN \eqref{eq:GN_update} by
\begin{equation}\label{eq:learnedGNStep}
p_k\left(x^{(k)},h\right)=(J_k^{\rm T}W_eJ_k + H_\Psi)^{-1}\nabla{\mathcal{E}\left(x^{(k)},h\right)}.
\end{equation}
Since the learned updates \eqref{eq:learned_iterative_update} harness the spatial information of the optical parameters, the classical regularizer $\Psi(x)$ is in practice only needed to stabilize the inversion in \eqref{eq:GN_update}. For this purpose, any differentiable regularizer, such as the weighted $l^2$-norm \eqref{eq:el_two} is suitable that allows using the update rule \eqref{eq:GN_update}. In case of a non-smooth regularizer such as TV, the learned minimization problem \eqref{eq:min_problem} can be formulated as a learned primal-dual problem~\cite{adler2018learned}. For the learned GD step \eqref{eq:learnedGDStep}, no regularizer was used to minimize the number of manually-set parameters. 

The gradient direction provides less information for the networks, but is significantly cheaper to compute and store than the GN step. During network training, the step direction is computed thousands of times, and using the GN step can cause a crucial slowdown. However, it is not clear if the gradients provide enough information for the networks to achieve feasible convergence within a small number $K$ of update layers $\Lambda_{\theta_k}$. Therefore, in hopes of reducing the computational burden while maintaining fast convergence, we introduce the learned quasi-Newton iteration scheme with
\begin{equation}\label{eq:LearnedQuasiStep}p_k\left(x^{(k)},h\right)=\bar{H}_k\nabla{\mathcal{E}\left(x^{(k)},h\right)},
\end{equation}
where $\bar{H}_k$ approximates the inverse of Hessian $H_k^{-1}$. The considered quasi-Newton approximation here is the symmetric-rank-1 (SR1) update 
\begin{equation}\label{eq:sr_update}
    \bar{H}_{k+1}=\bar{H}_k+\frac{\left(s_k- \bar{H}_k y_k\right)\left(s_k- \bar{H}_k y_k\right)^T}{\left(s_k- \bar{H}_k y_k\right)^T y_k},
\end{equation}
where $s_k=x^{(k+1)}-x^{(k)}$ and $y_k=\nabla{\mathcal{E}\left(x^{(k+1)}\right)}- \nabla{\mathcal{E}\left(x^{(k)}\right)}$. The initial approximation $\bar{H}_{0}$ is usually chosen to be a scaled identity matrix, leading to an initial step direction equal to the GD step. To avoid too small denominator values in \eqref{eq:sr_update}, a safeguard criteria such as~\cite{nocedal1999numerical}
\begin{equation*} 
    \left|y_k^T\left(s_k-H_k y_k\right)\right| \geq \omega\left\|y_k\right\|\left\|s_k-H_k y_k\right\|,
\end{equation*}
with a small $\omega>0$ can be applied to determine if the update needs to be skipped.

The main reason to choose the SR1 update over the popular BFGS update is its ability to produce stable Hessian approximations without an exact line-search~\cite{nocedal1999numerical}, which the learned iterative solvers cannot impose.

\subsection{Training schemes}
 Each of the learned iterative schemes \eqref{eq:learnedGDStep}-\eqref{eq:LearnedQuasiStep} consists of $K$ separate networks with respective weights $\theta_k$. Given supervised training data pairs $\{x_i,h_i\}_{i=1}^\mathcal{I}$ satisfying \eqref{eq:obs_model}, we aim to train the reconstruction operator $\mathcal{R}_\theta$ in \eqref{eq:recon_operator} defined by the updates \eqref{eq:learned_iterative_update} and where $\theta=\left\{\theta_0, \ldots, \theta_{K-1}\right\}$. The common ideal approach to train the reconstruction $\mathcal{R}_\theta$ operator is end-to-end, that means we train all iterates simultaneously by minimizing the following empirical loss 
\begin{equation*}
    \label{eq:etoeTraining}
    \theta^*=\arg\min_{\theta}\sum_{i=1}^\mathcal{I} \left\| \mathcal{R}_\theta(h_i) - x_i   \right\|_2^2.
\end{equation*}
In order to compute the backpropagation, all of the $K$ networks are unfolded. Therefore, each training iteration requires computing $K$ update directions and backpropagation through all intermediate steps. Evidently, training the networks in this end-to-end manner quickly becomes computationally infeasible if a large number of learned iterations is used or the dimensionality of the optical parameters is large. Additionally, the computation of the update directions needs to support either backpropagation or a have a defined derivative.

To avoid the need for backpropagation through the forward model and computations of the step direction, a greedy (iterate-wise) version of the optimization problem \eqref{eq:etoeTraining} can be considered. In the greedy training scheme each of the networks in \eqref{eq:learned_iterative_update} is trained separately and sequentially, forming a chain of $K$ training problems of the form 
\begin{equation}\label{eq:greedyTrain}
    \theta_k^*=\arg\min_{\theta}\sum_{i=1}^\mathcal{I} \left\| \Lambda_{\theta_k}\left(x^{(k)}_i, p_k\left(x^{(k)}_i,h_i\right)\right) - x_i   \right\|_2^2,
\end{equation}

\[
\begin{cases}
x_i^{(k)} &= \Lambda_{\theta_{k-1}^*}\left(x_i^{(k-1)}, p_{k-1}\left(x_i^{(k-1)},h\right)\right), \text{ for }k>0,\\
x_i^{(0)} & = [c_{\mu_a,i},c_{\mu'_s,i}].
\end{cases}
\]
In this work, we assume that for each sample there exists rough estimates $c_{\mu_a,t}\in\mathbb{R}$ for the magnitude of absorption and $c_{\mu'_s,t}\in\mathbb{R}$ reduced scattering for each sample and set uniform initial values $x_t^{(0)}=\left(\mathbf{1}c_{\mu_a,i},\mathbf{1}c_{\mu'_s,i}\right)$, where $\mathbf{1}\in\mathbb{R}^N$ is vector of ones. 

In the optical problem, we estimate reduced scattering and absorption simultaneously. The updating network $\Lambda_{\theta_k}$ then takes, absorption $\mu_a^{k}$, reduced scattering $\mu_s^{\prime(k)}$ and their respective step directions $p_k^{\mu_a}$ and $p_k^{\mu_s^\prime}$. However, if the scattering, absorption, and step directions are not correlated, forcing the networks to learn the correlations can hinder the performance. In this work, we consider networks $\Lambda_{\theta_k}^{\mu_s^\prime}$ and $\Lambda_{\theta_k}^{\mu_a}$ that separately update the absorption and reduced scattering based on their respective step directions.

In the end-to-end training \eqref{eq:etoeTraining}, the step directions $p_k$ are changing during the training and computed separately for each training iteration. Training the network End-to-end is expensive, as all intermediate steps of computing step directions are stored and backpropagated through. Whereas, in the greedy training \eqref{eq:greedyTrain}, training a single network $\Lambda_{\theta_{k}}$ requires computing the step directions only once for each sample, and the backpropagating is done merely through the network $\Lambda_{\theta_{k}}$. 

We emphasize that in this work, the update steps \eqref{eq:learnedGDStep}-\eqref{eq:LearnedQuasiStep} are computed with respect to the DA, but could be similarly implemented with other light propagation models. 

\subsection{Network architectures}\label{sec:archithecture}
\begin{figure}[h!tbp]
    \centering
    \includegraphics[width=1.0\linewidth]{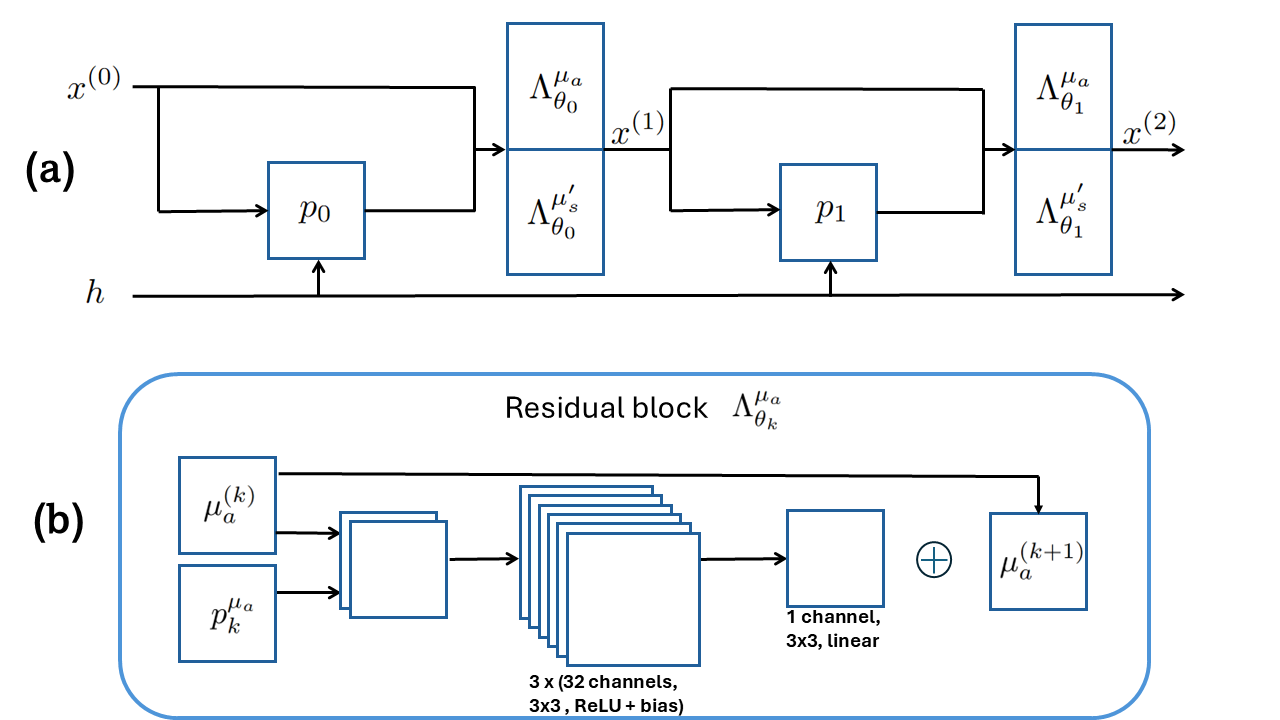}
    \caption{Used network architecture for the learned model-based iterative solvers. \textbf{a)} unenrolled first two iterations of the end-to-end trained iterative network that takes the initial optical values and the data $h$ as the input as in \eqref{eq:learned_iterative_update}. \textbf{b)} architecture of a single updating network $\Lambda_{\theta_k}^{\mu_a}$. The network takes the current absorption values $\mu_a^{(k)}$ and the respective step direction $p_k^{\mu_a}$ as two-channel input and uses 3 convolutional layers of 32 channels with ReLU, and a final single-channel layer with a final linear layer to produce the updated absorption values $\mu_a^{(k+1)}$. The networks $\Lambda_{\theta_k}^{\mu_s^\prime}$ in \textbf{a)} have the same architecture as shown in \textbf{b)}, but take the current reduced scattering $\mu_s^{\prime (k)}$ and respective step direction $p_k^{\mu_s^\prime}$ as the inputs.}
    \label{fig:flow}
\end{figure}

 The used network architecture $\Lambda_{\theta_k}$ for the model-based learned iterative methods followed the structure of the ResNet as earlier utilized in~\cite{mozumder2021model}. A single updating network block can be seen in Fig.~\ref{fig:flow}. The scattering and absorption were updated using separate networks, i.e, the two-channel input of one network was the current absorption $\mu_a^{(k)}$ and respective step direction, while the other network was given the current reduced scattering $\mu_s^{\prime(k)}$ and respective step direction. For both networks, the input was followed by three 32-channel convolutional layers with $3\times3$ kernel, bias, group norm of 8 groups, and ReLU activation. The fourth convolutional layer reduces the 32 channels to a single channel, which updates the current optical values. The last layer is not followed by ReLU, to allow negative optical parameter changes. To strictly restrict the positivity of the optical parameters, a ReLU activation function with a small additive bias of $10^{-5}$ was used at the end of each network $\Lambda_{\theta_k}$. 

To compare the model-based approach to a fully learned inversion scheme, a residual U-Net-based realization was implemented as in \eqref{eq:baselineUnet}. The input for the U-Net was the data, absorbed energy density, from which it directly produced the absorbtion and scattering. Again, two separate networks were used to produce absorption and scattering independently. The used residual U-Net follows a standard architecture with three scales (32, 64, and 128 channels), i.e., two max-pooling layers, and two convolutional layers on each scale for the encoding and decoding path.

\section{Experimental setups: data generation and implementation }\label{sec:simulations}

\subsection{Ideal simulated data}
 In the first experiment, an ideal setup was constructed by simulating a comprehensive amount of samples, in a highly diffusive region where the overall modeling error from using DA was negligible. This setup is ideal for comparing the convergence and reconstruction accuracy of the learned iterative methods with different step directions and training schemes.

For this experiment, we consider a rectangular 2-D domain (20\,mm $\times$ 25\,mm) for the optical parameters. The training samples were chosen to consist of randomly located and sized ellipsoidal inclusions, which were also allowed to overlap. The background optical values (mm$^{-1}$) of the samples were drawn to be uniformly between 0.0085 and 0.0115 for absorption and between 1.7 and 2.6 for reduced scattering. The contrast of absorption and scattering inclusions was uniformly drawn to be between 0.2 and 3.5 times the background optical values of each sample. The nodes were additionally corrupted by Gaussian noise with a standard deviation of 2\% of the amplitude. Finally, the samples were slightly blurred with a Gaussian filter to make the inclusion edges smoother, which would likely be the case with practical targets. These generated samples with ellipses can be interpreted to mimic, for instance, a cross-section of veins. 

\begin{figure}[h!tbp]
\centering
\includegraphics[width=0.47\textwidth]{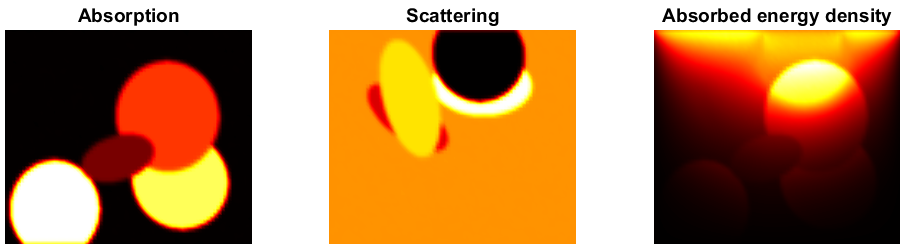}
\caption{\textit{Ideal problem}: Example of drawn absorption (left) and scattering (mid) coefficients and respective simulated absorbed energy density (right), when the top side was illuminated.}
\label{fig:ideal_example}
\end{figure}
To simulate the absorbed energy densities of each sample, the ValoMC~\cite{leino2019valomc} open Monte Carlo software package for Matlab was used. ValoMC utilizes the photon package method to simulate the fluence in the given geometry and light sources. The absorbed energy densities are then obtained by multiplying the fluence by the absorption as in \eqref{eq:E_density}. To guarantee a unique solution when estimating scattering and absorption simultaneously, two separate limited-angle illuminations were performed, forming two sets of absorbed energy densities for each sample. The illuminations were simulated from the top and right side of the rectangle using $10^8$ photons uniformly entering from the sides. By default, the ValoMC scales the total strength of the light source to correspond to 1\,W. Example of drawn absorption and scattering coefficients and respective simulated absorbed energy density (top side illuminated) are shown in Fig.~\ref{fig:ideal_example}. The simulated absorbed energy density was corrupted by noise drawn from a Gaussian distribution with a standard deviation corresponding to 1\% of the absorbed energy density values. The simulations were repeated with 1250 samples.

\subsection{Digital twin phantoms}\label{sec:digital_twin_sim}
\begin{figure}[h!tbp]
    \centering
    \includegraphics[width=0.45\textwidth]{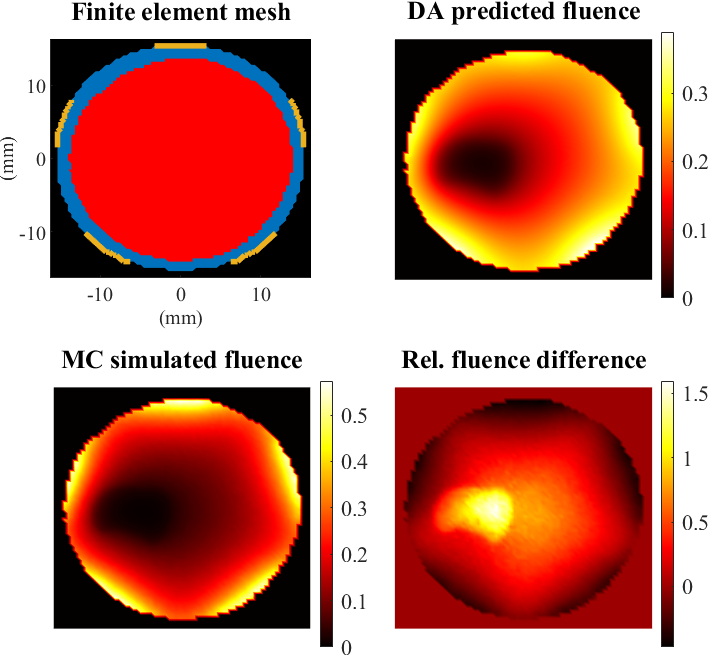}
    \caption{\textit{Digital twin problem}: Used finite element mesh for the digital twin problem (left-up). The blue regions represent the water nodes, red the estimated nodes, and yellow the light source locations. The right-top figure shows an example of predicted fluence $\Phi_\text{DA}$ from a 2D diffusion approximation for a \emph{in-dist.} sample. The bottom-left figure shows the respective fluence $\Phi_\text{sim}$ from a 3D Monte Carlo simulation averaged over the out-of-plane dimension. The bottom-right figure shows the node-wise relative difference $(\Phi_\text{DA} - \Phi_\text{sim})/\Phi_\text{sim}$.}
    \label{fig:mod_error}
\end{figure}

 We refer to the second numerical example as  the digital twin problem. It is much closer to experimental conditions compared to the ideal problem. The problem now has limited training data, samples with varying locations of the tubes (region of interest), a large range of possible optical values, and a large modeling error (see Fig.~\ref{fig:mod_error}).

We emphasize that since the data is statistically simulated based on physical absorption and scattering phantoms, this corresponds to a case where the acoustical problem is perfectly solved, and possible noise from the optical inversion is not apparent.

The digital twin problem re-used data from physical tissue-mimicking phantoms with piecewise-constant material distributions~\cite{grohl2023moving,grohl2023dataset}, enabling optical characterization via an in-house double integrating sphere system~\cite{hacker2021copolymer,pickering1993double} to determine wavelength-dependent absorption ($\mu_a$) and reduced scattering ($\mu_s'$) coefficients between $600 \text{ and } 950\,\mathrm{nm}$. The phantoms were immersed in a water bath and imaged using a pre-clinical PAI system (MSOT inVision-256TF, iThera Medical GmbH, Munich, Germany). It has 256 transducer elements in a circular array with an angular coverage of 270$^\circ$, which have a 5-MHz centre frequency and 60\% bandwidth. We imaged in the wavelength range from 700\,nm to 900\,nm in 10\,nm steps using 10-frame averaging.

We constructed digital twins of the physical phantoms using manual segmentation masks derived with MITK~\cite{nolden2013medical} based on images obtained with filtered backprojection. We assigned our reference optical measurements and reference acoustic characterisation from the literature (density: $1000\,\mathrm{g/cm}^3$; sound speed: $1468\,\mathrm{m/s}$ for phantoms, $1489\,\mathrm{m/s}$ for coupling medium) to each piecewise-constant material region. 

Our simulation pipeline computes device-specific initial pressure distributions $p_0(r)$ and measurement data $p(t, r)$ as follows:
\begin{enumerate}
    \item Light fluence $\phi$ was simulated using MCX~\cite{fang2009monte}, yielding $p_0(r) = \Gamma \cdot \mu_a(r) \cdot \phi(r)$ with constant Gr\"uneisen parameter $\Gamma$. 

    \item Acoustic wave propagation was modeled using k-Wave's k-space pseudospectral method~\cite{treeby2010k} to generate $p(t, r)$.
\end{enumerate}
The generated time-serieses $p(t,r)$ were only used to derive the segmentation and not used to obtain the reconstructions in section \ref{sec:results}. Instead, with the constant Gr\"uneisen parameter, the used data for the optical inversion was $h(r)=\frac{p_0(r)}{\Gamma}$.

A digital twin of the MSOT InVision-256TF implemented in SIMPA~\cite{grohl2022simpa} incorporated vendor-provided hardware geometry. Custom MCX illumination and k-Wave transducer arrays were validated against $15$ calibration phantoms. Optimized Gaussian illumination profiles achieved experimental agreement in radiant exposure measurements. SIMPA orchestrated all simulations through integrated MCX and k-Wave adapters. More detailed information on the digital twin construction can be found in~\cite{grohl2025digital}, details on the phantoms in~\cite{hacker2021copolymer}, and details on the imaging system in~\cite{grohl2023moving}.

The total number of actual digital twin samples for training and testing was only 17. As such, we supplemented the digital twins with additional phantom simulations, where the optical properties are not in correspondence with physical phantoms but instead determined the following way: (1) Draw two random material characterizations from the physical phantom data, $M_1$, $M_2$; (2) Determine a random mixing factor $c=[0,1)$; (3) Calculate the new material property $M_3 = c M_1 + (1-c) M_2$. The resulting distribution can be found in the supplementary file Fig.~S1. Each phantom was drawn to have between 1 and 3 inclusions and a radius of 14.2\,mm. We simulated at six wavelengths: 700, 740, 780, 820, 860, and 900\,nm. In total, we supplemented the training data with 41 of these phantom simulations.

The digital twin problem introduces several sources of modeling error. Firstly, the simulations were conducted in 3D, but we approximated the light distribution in 2D, where the photons need to travel a much shorter path to reach the center, causing underestimated decay of fluence towards the center as in Fig.~\ref{fig:mod_error}. Secondly, the optical values of the samples contained relatively high absorption coefficients, violating the $\mu_a << \mu_s'$ assumption of the DA. hence causing it to be inaccurate. An example of this inaccuracy can be seen in Fig.~\ref{fig:mod_error}, where the DA clearly predicts too spread fluence field, compared to the MC simulated fluence.

\subsection{Implementing the forward models}\label{sec:implementation}

 The FE solution of the diffusion approximation was implemented according to the equations in Sec.~\ref{sec:forward} using PyTorch tensors and built-in functions. A piecewise constant basis was used for the optical parameters. In the case of the ideal simulated data, the whole 2-D square domain was discretized with 4636 nodes, forming 9000 evenly sized triangular elements. For digital twin discretization, most of the water around the tubes was left out of the FE discretization. However, as the same FE mesh was used for all of the samples and the location of the tubes slightly varied, a small layer of water remains in the discretization region, as demonstrated in Fig.~\ref{fig:mod_error}. 

For the convolutional networks, the discretization was embedded back to an evenly spaced grid of $72\times72$ nodes. The FE discretized part of the domain had evenly spaced 4172 nodes and 8148 evenly sized triangular elements. The location of the tubes in the water was assumed to be known, and only the non-water nodes ($\sim 90\%$ of the mesh) were estimated. 

For the ideal simulated data, DA was implemented with light sources that uniformly illuminated the top and right sides, similar to the ValoMC simulation. Whereas, for the digital twin experiments, the five 6.12\, mm wide light sources were approximated to have a Gaussian intensity profile with a standard deviation of 3. 

\subsection{Backpropagation and computational considerations}\label{sec:consider}
 For the End-to-End training \eqref{eq:etoeTraining}, computation of the update steps \eqref{eq:learnedGDStep}-\eqref{eq:LearnedQuasiStep} is implemented inside the networks. This includes computing the FEM matrices \eqref{eq:fluence}, the corresponding fluence \eqref{eq:fluence}, and solving the chosen step direction \eqref{eq:learnedGDStep}-\eqref{eq:LearnedQuasiStep}. As these computations come with a large number of operations for which gradients are needed with respect to the weights of the networks, it is convenient to utilise the automatized gradient computation provided by PyTorch. PyTorch currently supports GPU versions for all the required matrix operations to compute the fluence and step directions. The detailed steps to compute the Jacobian of \eqref{eq:min_problem} with respect to scattering and absorption can be found, for instance, in~\cite{tarvainen2013image}.  

For each of the step directions, the weighting matrix $L_e$ in \eqref{eq:min_problem} needed to be determined. The noise profile for the ideal simulations, was known to be Gaussian $e\sim \mathcal{N}(0,\Gamma_e)$ (with 1\% standard deviation of the data amplitude). Therefore, the weight matrix $L_e$ was chosen to be the Cholesky factor of the inverse of the covariance $\Gamma_e^{-1}=L_e^TL_e$. The noise profile of the digital twin was unknown, and the weighting matrix $L_e$ was set as the identity matrix. 

In the digital twin simulations, the samples had an extensive range of data values, and weighting each node equivalently led to poor convergence of certain regions. To overcome this, a log-scaled data space was used, substantially improving the performance of both classical and the learned solvers when used for the digital twin problem. Using log-scaled space for the ideal problem that had a narrow range of data values did not show a noticeable difference in performance.  

Solving the digital twin problem uniquely required assuming the wavelength dependency, for the scattering. The training samples were fit to approximately follow $\mu_s^\prime(r,\lambda)\approx a(r)e^{-b\lambda}$, dependency with sample-specific constant $b>0$. However, the constants $b$ were varying only slightly, and using the average value $1.6\times10^{-3}$ was observed to work sufficiently for the training and test samples.

\subsection{Training procedures and computational requirements}
 The training-validation-test splits were done as follows. For ideal simulations, a total of 1250 samples were drawn, with a training-test-validation split of 80\,:10\,:10. For digital twin experiments, the samples were roughly labeled as follows: \textit{i)} The generated complementary \emph{augmented} samples that had absorption and scattering drawn as described in Section \ref{sec:digital_twin_sim}, \textit{ii)} in-distribution (\emph{in-dist.}) digital twin samples that represent the experimental phantoms and are similar to \emph{augmented} samples, \textit{iii)} \emph{outlier} digital twin samples that represent the experimental phantoms with much larger inclusions. The training set consisted of 7 \emph{in-dist.} samples and 41 \emph{augmented} samples. Since the illumination patterns were approximately symmetric with respect to the vertical axis (Fig.~\ref{fig:mod_error}), the training set was augmented by flipping the vertical coordinates of the samples, doubling the size of the training set. The validation set consisted of 4 \emph{augmented} targets, and the test set 5 \emph{in-dist.} samples, 6 complementary \emph{augmented} samples, and 5 \emph{outlier} samples. Each sample was solved with six different wavelengths, 700, 740, 780, 820, 860, 900 (nm), yielding a total of 6$\times$98 training samples and 6$\times$17 test samples. However, the optical values and locations of the inclusion were highly correlated and only marginally different for the six wavelengths used. 

For all training tasks, the $\ell^2$ loss function was used. Due to reduced scattering being significantly larger than absorption, in the loss function, the scattering of digital twin samples was weighted by a factor of 1/10, and by a factor of 1/100 for ideal samples, balancing the loss contribution. The model-based iterative learned solvers were trained using 35000 (ideal) and 40000 (digital twin) training steps. The learning rate followed cosine annealing scheduling with initial learning rates of $10^{-4}$ (ideal) and $4\times10^{-5}$ (digital twin). Due to the volatile behaviour of the optical values inside untrained end-to-end networks $\Lambda_{\theta_k}$, a few thousand smaller initial learning steps were used to avoid numerical errors and unstable inversion. 

The memory consumption of the greedy training was primarily contributed by the stored Jacobian and Hessian, yielding less than 2 GB of peak GPU memory usage for all greedily trained solvers. For end-to-end training, all of the intermediate steps of computing the fluence and step direction needed to be stored for efficient backpropagation. This produced a linear increase in the memory consumption with respect to the number of used updating networks $\Lambda_{\theta_k}$. The memory consumption increases of a single updating network $\Lambda_{\theta_k}$ within the ideal problem are shown in Table~\ref{tab:run_times}.

\begin{table}[h!tb]
\centering
\caption{
Memory consumption and run times}
\label{tab:run_times}
\begin{tabular}{lllllll} 
\hline
\multirow{2}{*}{Solver} & \multicolumn{2}{l}{$t_{forw}$ (s)} & \multicolumn{2}{l}{$t_{back}$ (s)} & \multicolumn{2}{l}{Memory (GB)} \\ 
\cline{2-7}
  & Ideal  &  Twin  & Ideal  & Twin & Ideal  & Twin  \\ 
\hline
GD   & 0.32  &  0.34  &  0.51  &  0.23    & 2.7  &  1.8                                 \\ 

GN & 0.62  &  0.56 &  0.82  &  0.36 & 3.4   &  3.1                              \\ 
SR1  &   0.33  &  X   &  0.52  & X & 3.4 & X                       \\
U-Net  &    0.01 &  $<$\,0.01   &   0.03  &  0.01  & 0.1  &  0.1
                      \\
\hline
\end{tabular}
\end{table}

In the greedy training, the majority of the training time was spent on solving the fluences and step directions for each sample. The time to compute forward pass and backpropagation through only an updating network $\Lambda_{\theta_k}$ took less than 0.01\,s. Therefore, the total training time for the greedy training tasks was approximately $S \times t_{forw} \times K$, where $S$ is the number of training samples, $K$ is the number of updating networks $\Lambda_{\theta_k}$, and $t_{forw}$ is the time of computing a single step direction. The times $t_{forw}$ to compute the different step directions are shown in Table~\ref{tab:run_times}. For instance, five updating networks of GN step greedily trained with 35000 iterations using 1000 samples, took approximately one hour.

In the end-to-end scheme, evaluating the forward pass took roughly the same amount $t_{forw}$ as in the greedy scheme, but now the backpropagation was done through all of the intermediate steps of computing the step directions. Also, since the step directions (except the initial step directions) were changing during the training, computing the $K$ step directions and back propagation through them was repeated for each training iteration $T$. Therefore, the total training time for the end-to-end tasks was approximately
$T \times (t_{forw}+t_{back}) \times (K-1)$, where $t_{back}$ is the time taken to backpropagate through a single updating network and the intermediate steps of a single step direction computation. The average forward pass and backpropagation evaluation times for the end-to-end solvers are shown in Table~\ref{tab:run_times}. For instance, within an ideal problem, five end-to-end trained updating networks with GN step and 35000 training iterations took roughly 62 hours to train. 

The residual U-net used for the fully learned approach took approximately 30 minutes to train with 80000 training iterations. The reconstruction (forward pass) was evaluated in less than 0.01\,s. All of the computations were performed using an RTX A5000 GPU equipped with 24GB of memory. All of the used codes can be found on GitHub (\url{https://github.com/AnsManni/Learned_iterative_QPAT})

\section{Results}\label{sec:results}

\subsection{Ideal problem}
 In the first study, we tested greedy and end-to-end trained GD, GN, and SR1 update-based learned iterative methods on the ideal simulated problem. To compare the performance of using varying numbers of updating networks $\Lambda_{\theta_k}$, multiple learned iterative solvers were trained, using between 1 and 9 updating networks. As a comparison, a residual U-Net-based fully learned reconstruction method was tested. Further, a single test sample was reconstructed using the classical GN updates with $\ell^2$ regularizer utilizing local spatial correlations via the Ornstein-Uhlenbeck process \eqref{eq:O-U} with correlation length of 1\,mm and standard deviation of 0.005 for absorption and 0.035 for scattering. 
\begin{figure}[b!]
    \centering
    \vspace{-0.4cm}
    \includegraphics[width=0.345\textwidth]{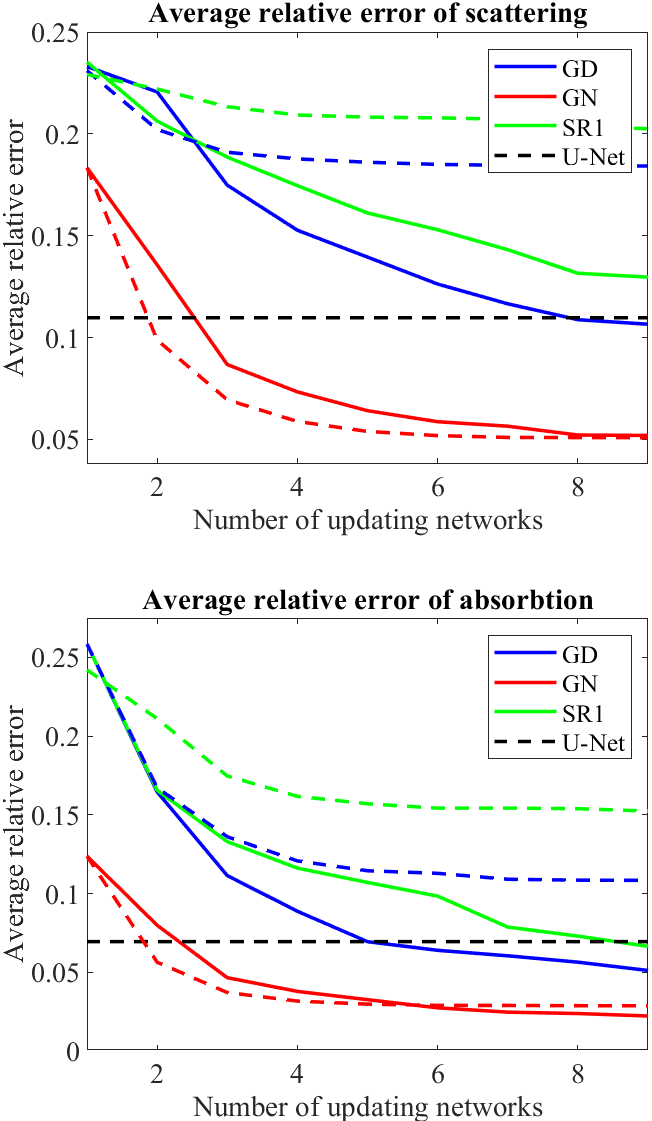}
    \caption{\textit{Ideal problem}: Average relative errors of absorption ($\mu_a$) and reduced scattering ($\mu_s^\prime$) over 125 test samples. The solid lines show the relative errors of end-to-end trained learned iterative methods up to 9 updating networks $\Lambda_{\theta_k}$ and the dashed lines the corresponding greedily trained values. The black dashed line shows the average relative error from the residual U-Net.}
    \label{fig:relative_errors_ideal}
\end{figure}
The learned GN used $\ell^2$-regularizer with regularization parameters of 10000 for absorption and 1.8 for scattering. All initial values for the classical and learned solvers were set to correspond to the average absorption (0.01) and scattering (2) values over the training set, which were also used as the mean of the $\ell^2$-regularizer \eqref{eq:el_two}.

For the classical GN updates, we found a bisection style inexact line-search~\cite{line_search} to be efficient. For the learned GN updates, we used an uncorrelated $\ell^2$-regularizer \eqref{eq:el_two} with regularization parameters of 12.5 for absorption and 0.17 for scattering. 

In general, the performance of the classical solver was heavily affected by the chosen regularizer and regularization parameters. In our experiments, we found that including local spatial correlations in the form of the Ornstein-Uhlenbeck process produced stable reconstructions across the simulated samples. On the other hand, the performance of the learned GN and SR1 was consistent over a wide set of regularization parameters and only started to decrease once the regularization parameters were drastically increased.

\begin{figure}[h!tbp]
\centering
\vspace{-0.3cm}
\includegraphics[width=0.5\textwidth]{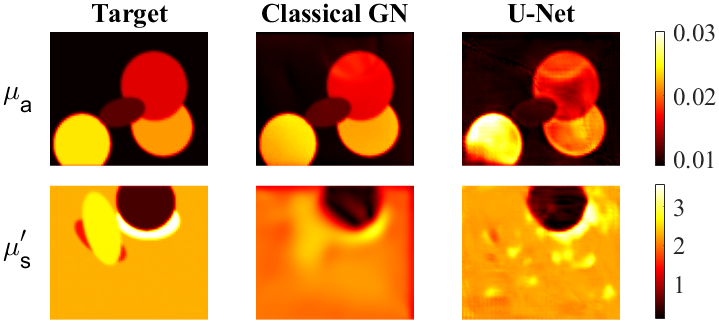}
\caption{\textit{Ideal problem}: absorption ($\mu_a$) and reduced scattering ($\mu_s^\prime$) reconstructions of a test sample (left) using classical GN solver (14 iterations) and residual U-Net.}
\label{fig:recos_ideal_target}
\end{figure}

\begin{figure}[h!tbp]
\centering
\includegraphics[width=0.5\textwidth]{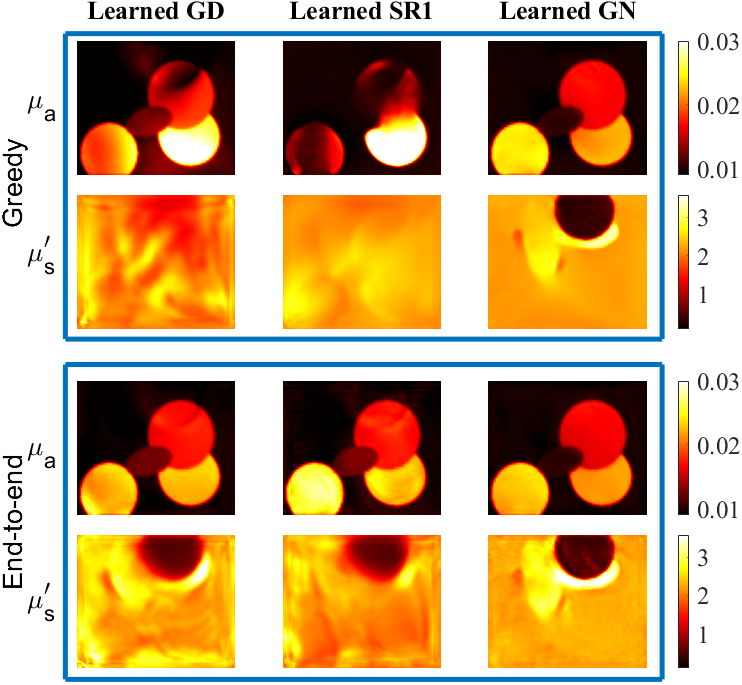}
\caption{\textit{Ideal problem}: Absorption ($\mu_a$) and reduced scattering ($\mu_s^\prime$) reconstructions of a single test sample using greedily (top) and end-to-end (bottom) trained learned iterative SR1, GD, and GN with 9 updating networks $\Lambda_{\theta_k}$. }
\label{fig:recos_ideal_learned}
\end{figure}

Fig.~\ref{fig:relative_errors_ideal} shows the average relative error of absorption and reduced scattering over the test set of 125 samples for GD, SR1, and GN based learned iterative methods up to 9 updating networks $\Lambda_{\theta_k}$. After 9 iterations, the difference in average relative errors was observed to be marginal. The average relative error of the residual U-Net reconstructions is shown as the horizontal dashed line in Fig.~\ref{fig:relative_errors_ideal}. Reconstructions of a single test sample are shown in Fig.~\ref{fig:recos_ideal_target} for the classical GN solver and the residual U-Net and in Fig.~\ref{fig:recos_ideal_learned} for each of the greedily and end-to-end trained methods.

Relative errors in Fig.~\ref{fig:relative_errors_ideal} reveal the learned GN solver having significantly better performance compared to the GD and SR1. The learned GN based solver is also the only one to noticeably surpass the accuracy of the residual U-Net. For the learned GN updates, there was no distinguishable difference between the end-to-end and greedily trained networks. On the other hand, the learned GD and SR1 yielded relatively accurate absorption and scattering reconstructions, but only when trained end-to-end. The generally poor performance of the greedily learned iterative solvers can be seen in Fig.~\ref{fig:recos_ideal_learned}. Only the GN updates provided sufficient information to the networks making the greedy training produce feasible absorption and scattering reconstructions (Fig.~\ref{fig:recos_ideal_learned}). 

The residual U-Net was also able to produce accurate reconstructions with an average relative error of 7\% for absorption and 9\% for reduced scattering. As can be seen from Fig.~\ref{fig:recos_ideal_target}, despite low relative scattering error, the residual U-Net tends to produce many hallucinated artifacts in the scattering reconstructions. The learned iterative methods did not in general, produce hallucinated scattering artifacts (Fig.~\ref{fig:recos_ideal_learned}).

The reconstructions with the classical GN in Fig.~\ref{fig:recos_ideal_target} demonstrate that with relatively low modeling error, even the classical solver, fine-tuned with proper regularization parameters, can accurately reconstruct both optical parameters. However, compared to the learned solvers, the classical solver required more GN iterations, each requiring multiple forward model evaluations to determine a suitable step length. 

\subsection{Digital twin problem}\label{sec:DigTwin}
 In the digital twin study, we reconstructed a set of digital phantoms with approximately the same shapes and optical parameters as existing experimental phantoms. We used a finite element approximation of the DA light propagation model based on the real measurement setup. The reconstruction was done using end-to-end GD and GN based learned iterative methods, same residual U-Net-based architecture as in the ideal study, and a classical GN solver with the total variation regularizer \eqref{TV_seminorm}, ideal for reconstructing the piecewise constant inclusions. Up to five end-to-end trained GD/GN updating networks were used, as no notable improvement was achieved with further networks. The learned SR1 was excluded from this study due to its generally poor performance that was observed outside of the reported results.

The initial values $x^{(0)}$ were drawn to be uniformly 0-30\% off from the ground truth background optical values of each sample. These initial values were also used as the mean $x_\mu$ of the $\ell^2$ regularizer \eqref{eq:el_two} for the learned GN. The classical total variation regularized problem was solved by using the primal-dual interior point (PD-IPM) method~\cite{andersen2000efficient}. The total variation regularizer was used with a smoothness parameter of $10^{-4}$ and a regularization parameter of 0.1.

\begin{figure*}[tbp!]
    \centering 
    \includegraphics[width=0.9\textwidth]{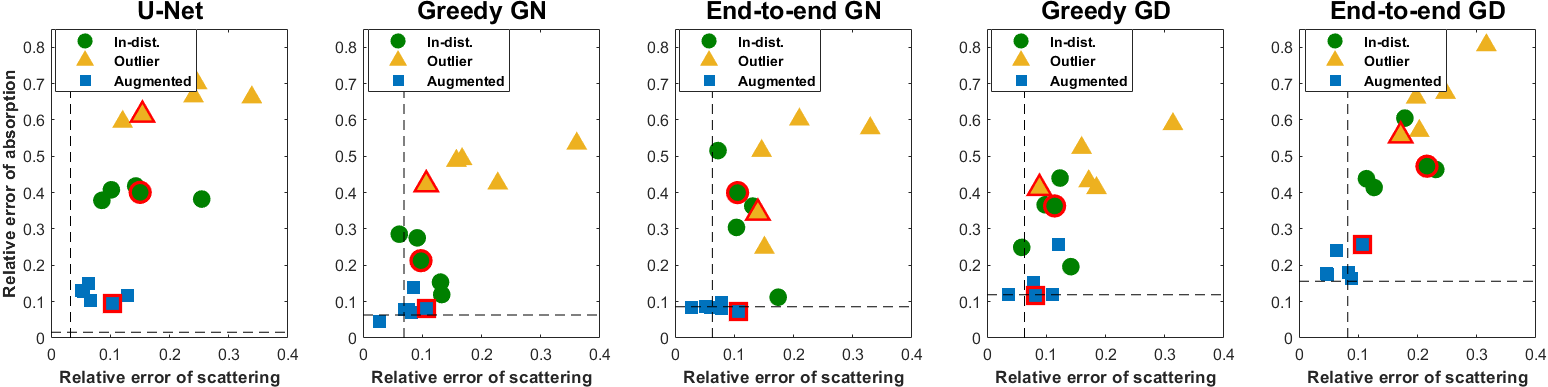}
    \caption{\textit{Digital twin problem}: relative errors of absorption ($\mu_a$) and reduced scattering ($\mu_s^\prime$) for \emph{augmented}, \emph{in-dist.}, and \emph{outlier} test samples. Starting from left, the relative errors are from the residual U-Net, greedily trained iterative Gauss-Newton, end-to-end trained Gauss-Newton, greedily trained iterative Gradient descent, and end-to-end trained Gradient descent. The dashed black line shows the average relative error of training samples. The relative error of each plotted sample is averaged over the six solved wavelengths. The learned iterative methods use five updating networks $\Lambda_{\theta_k}$. The highlighted samples correspond to the shown samples in Fig.~\ref{fig:recos_twin}.}
    \label{fig:relative_errors_twin}
\end{figure*}

Fig.~\ref{fig:relative_errors_twin} shows the relative errors of each reconstructed test sample type using residual U-Net, learned iterative GD, and GN, both with 5 updating networks. The relative errors show that all of the learned methods performed well with the \emph{augmented} test samples. 

From Fig.~\ref{fig:relative_errors_twin} we can see both learned iterative GN based solvers performing most robustly over all of the sample types. The residual U-Net performs at a similar level of accuracy with the \emph{augmented} samples, but considerably drops in reconstruction quality for \emph{in-dist.} and \emph{outlier}. The most challenging \emph{outlier} samples were, in general, difficult to reconstruct for all learned methods due to large inclusions with high absorption, producing a completely different fluence field compared to training samples.

For residual U-Net, the amount of training samples was evidently too low, as the accuracy difference between the training dashed lines and test reconstructions was significant, as shown in Fig.~\ref{fig:relative_errors_twin}. The learned iterative solvers generalized well with the \emph{augmented} test samples, but less well with the \emph{in-dist.} and \emph{outlier} samples. 

\begin{figure*}[tbp!]
    \centering
    \includegraphics[width=1\textwidth]{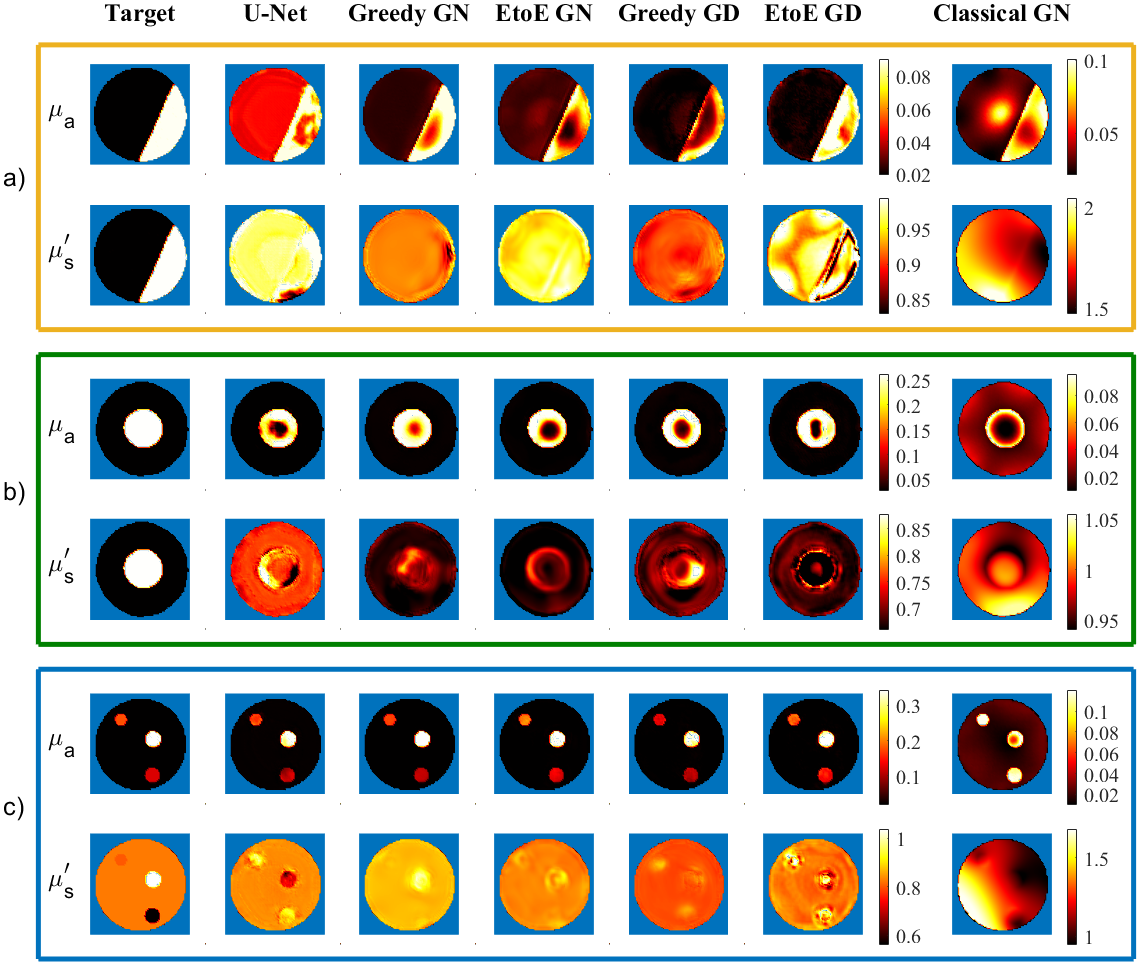}
    \caption{\textit{Digital twin problem}: absorption ($\mu_a$) and reduced scattering ($\mu^\prime_{s}$) reconstructions of \textbf{a)} \emph{outlier}, \textbf{b)} \emph{in-dist.}, and \textbf{c)} \emph{augmented} test samples (leftmost column) when the wavelength of the light sources was 780 nm. The methods used for the reconstructions starting from the second leftmost column are residual U-Net, greedily trained iterative Gauss-Newton, end-to-end trained Gauss-Newton, greedily trained iterative Gradient descent, end-to-end (EtoE) trained Gradient descent, and classical Gauss-Newton with total variation regularizer after 28 iterations. The learned iterative methods were using five updating networks $\Lambda_{\theta_k}$. The reconstructions from the classical Gauss-Newton are shown with a separate color scale from the others. A small percentage of the lowest and highest valued nodes of the learned reconstructions are excluded from the color scale.}
    \label{fig:recos_twin}
\end{figure*}

Fig.~\ref{fig:recos_twin} shows example reconstructions with classical solver and the learned methods, for all sample types. For the classical solver, the modeling error was overwhelming, causing a very poor reconstruction quality for both scattering and absorption as seen from Fig.~\ref{fig:recos_twin}. We tested the classical solver with different regularization and smoothing parameters, but none of the changes generally improved the reconstruction quality.  

\section{Discussion}\label{sec:discussion}
\subsection{Ideal problem}
 In the first study, we investigated the performance of different learned iterative model-based solvers when used to solve a nonlinear inverse problem under ideal conditions. The results showed significant differences in reconstruction accuracy when the GD, GN, and SR1 step directions were used as the information for the networks. Increasing the number of trained networks increased performance, analogous to the behavior of classical iterative solvers. However, all the learned iterative solvers except the greedily trained SR1, were drastically more robust than the classical GN solver. The performance of the learned solvers was not sensitive to the chosen prior values, whereas finding a plausible prior value for the classical solver was not trivial.

In general, the learned GN achieved superior reconstruction accuracy compared to the other methods. However, the whole Hessian was computed for GN updates, leading to larger memory consumption and longer training times. While the aim of utilizing the learned SR1 update was to match the convergence speed and accuracy of the GN updates, our study showed that the SR1 step direction seems to provide information similar to the GD step. However, the expensive matrix inversion needed for the GN step could also be approximated by using existing matrix-free iterative techniques.

For a much larger dimensional problem, the training procedure (scheduling, initial weights, etc.) should be optimized as much as possible to reduce training time. Also, the resolution of the first few networks could be coarser and refined in later iterations to reveal finer targets. It could also be beneficial to set the initial optical values more accurately, based on, for instance, a few greedy iterations or the output of the residual U-Net. 

\subsection{Digital twin}
 In the second study, we investigated the performance of the implemented solvers in a more challenging setup, incorporating only a small training set and a large modeling error. In contrast to the ideal problem, the classical GN solver was now generally unable to reconstruct the targets due to the modeling error (see Figs~\ref{fig:hook} and \ref{fig:recos_twin}). On the other hand, the fully learned solver, which was based on the residual U-Net architecture, performed comparably well to the learned iterative methods. The good performance is not unexpected, as the input to the residual U-Net remains to be the absorbed energy density with only little noise and as long as test samples are consistent with the training the U-Net generalizes well.

Comparing greedily and end-to-end trained relative errors from Fig.~\ref{fig:relative_errors_twin}, we see that the greedy variants produce slightly smaller relative errors compared to end-to-end variants. This smaller error has also been observed during training, even though in theory, end-to-end training should provide an upper bound for the training performance of the greedy training. However, the optimization problem of the end-to-end scheme is far more complex and highly non-linear.
In general, the optical values fluctuate more during the intermediate GN/GD updates in the end-to-end training compared to the greedy training. The varying optical values are likely to produce extreme values in the update directions, yielding less stable convergence of the training.

One question was whether the augmented \emph{augmented} training samples were good enough for the U-Net and iterative learned networks to generalize to the \emph{in-dist.} and \emph{outlier} samples. The relative errors in Fig.~\ref{fig:relative_errors_twin} demonstrate that in fact, the iterative networks did generalize quite well for the \emph{in-dist.} samples, but not for the \emph{outlier}. 
The U-Net achieved the highest accuracy on training samples, but this accuracy failed to generalize well to the test samples.

While the learned solvers steadily improved the reconstruction accuracy with respect to the number of updating networks in the ideal study, they were now not able to achieve similar convergence. The main obstacle to the performance of the iterative learned solvers was the modeling error, which was not directly compensated. The end-to-end networks indirectly compensate for the modeling error as the information of the gradients flows through the networks, whereas the greedy training computes erroneous step directions and is unable to improve after the first network. The step directions inside the end-to-end network are computed according to the term $\|h-F(x)\|^2_2$, corrupted by the modeling error. To make the gradients more informative, a more accurate light transport model could be considered. 

Using more accurate models, such as the RTE, comes with a heavy computational cost. However, combined with greedy training, it can still be faster than using less accurate models with end-to-end training. The question would be whether a greedily trained solver with a more accurate model performs better than an end-to-end trained solver with an inaccurate model. 

Instead of using more expensive models, a modeling error compensation technique could be considered, too. However, as traditional error compensation methods, such as Bayesian approximation error modeling, utilize a large number of samples to estimate error statistics, it is not clear if these methods can provide a remedy if only a low number of training samples is available. As the behavior of the modeling error depends on the optical values and their relative spatial distances to each other, using an additional convolution-based correction layer could allow for a rough estimation of the modeling error.

\subsection{Modeling error}
One central focus of the digital twin problem was to investigate how the learned iterative methods perform under prominent modeling errors. The results presented in Section \ref{sec:DigTwin} demonstrate that the iterative methods perform relatively well with test samples that have a similar optical value pattern and hence a modeling error pattern to that of the training samples. However, the relative errors of the \emph{in-dist.} and \emph{outlier} indicate that further model error compensation is needed to improve robustness with more out-of-distribution targets. In other words, this work investigated the capability of the networks to implicitly learn a model correction \cite{hauptmann2018approximate,mozumder2021model} from the training samples as part of the updates.

We will here shortly discuss a few possible ways to explicitly compensate for modeling errors. Firstly, we note that, as the modeling error distribution is generally unknown and the sample size is small, classical statistical methods such as the Bayesian approximation error modeling are not likely to work \cite{arridge2006approximation}. Alternatively, one could train a dedicated network to learn an explicit model correction \cite{lunz2021learned}. However, in our experience, these networks are likely to face similar problems with the out-of-distribution targets and with respect to sample size. The obvious solution to this is to simulate a large set of samples and extend the training set to be more versatile. In this work, we used a small set of additional simulated samples to augment the digital twin samples, to maintain an emphasis on the digital twins.

If one wants to design a problem specific model correction we need note that the modeling error emerges from two factors: 1) the high non-diffusivity of the estimated regions, 2) Data averaged from a 3D forward model is solved with the 2D diffusion approximation. The 3D-to-2D modeling error increases with respect to the estimation depth due to incorrect photon fluence decay rate. Assuming approximately constant optical parameters along the non-plane axis would enable compensating the 3D-to-2D error separately. The decay rate depends on the magnitude of the absorption and scattering. Learning a depth-wise correction for the photon fluence could be key for compensating for the 3D-to-2D modeling error. 
Compensating for the discussed modeling errors becomes increasingly important when we shift from the simulated data to experimental data, where one also needs to solve the acoustical problem first. Nevertheless, designing such problem specific model corrections is beyond the scope of this work.

\begin{figure}[h!tbp]
    \centering
    \includegraphics[width=0.47\textwidth]{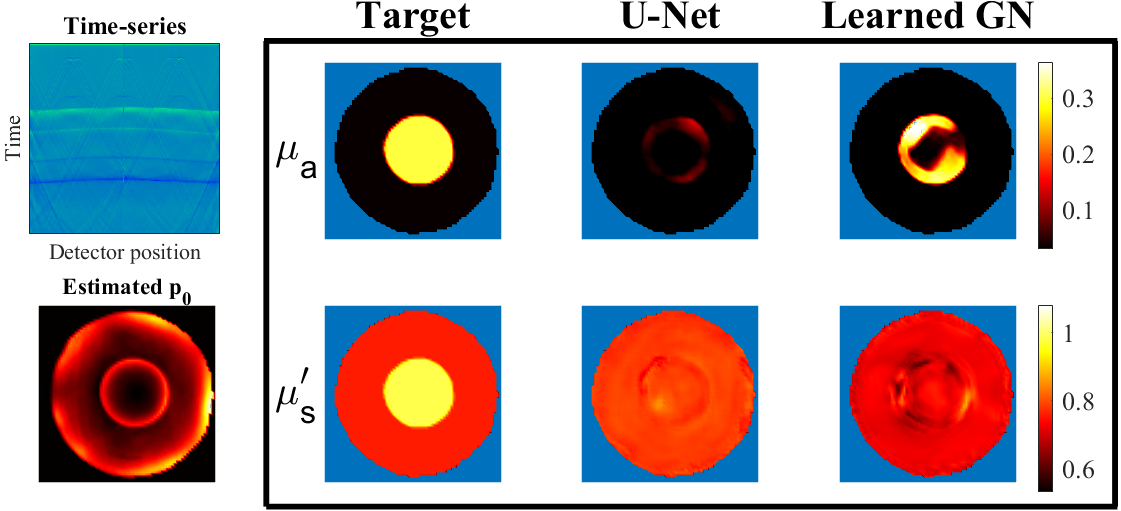}
    \caption{
    Absorption ($\mu_a$) and reduced scattering ($\mu^\prime_{s}$) reconstructions of a test sample solved from experimental time-series data, using residual U-Net and greedily trained iterative Gauss-Newton (Learned GN). The left-most column shows an initial acoustic field $p_0$ (bottom) that is estimated from the shown time-series (top) using a classical iterative solver.}
    \label{fig:experimental}
\end{figure}

\subsection{Robustness under experimental data}
This paper focused on solving the optical problem robustly under the assumption that the acoustical problem is solved almost perfectly. Therefore, we emphasize that simply applying the trained learned iterative methods or U-Net to reconstructions from the acoustic problem will not be successful.

Nevertheless, first results from the corresponding experimental data for the digital twins can be obtained when the networks are retrained on classical reconstructions from the acoustic problem. For this purpose we considered the time-series measurements for the physical phantoms corresponding to the digital twin samples (7 for training) together with the same 41 \emph{augmented} samples previously used. Reconstructions of the acoustic problem have been obtained from the corresponding variational problem with TV regularization solved using primal-dual hybrid gradient, as discussed in \cite{hauptmann2025fast}. To compute the inverse, adjoint, and forward acoustical operators required for the initial estimate and gradient update, we utilized a fast FFT-based solver \cite{hauptmann2025fast}. We have then trained the residual U-Net and the greedy learned GN following the same training procedure as earlier. The Grüneisen parameter was assumed to be constant ($\Gamma = 1$), and therefore, the absorbed energy densities were directly obtained from the acoustical inversions.

Reconstructions of the same test sample as in Fig.~\ref{fig:recos_twin} \textbf{b)} are shown in Fig.~\ref{fig:experimental}. Additional reconstructions of other test samples, can be found in Section D of the supplementary file.
The residual U-net now produces many artefacts in the background (this can be seen better with a more compressed color scale) and is unable to reconstruct inclusions in the deeper regions. Whereas the learned GN manages to reproduce the inclusions more accurately. Nevertheless, these reconstructions demonstrate that the estimation in the deeper regions, where the acoustic reconstruction and modeling error are most inaccurate, needs to be improved.

This result illustrates that the proposed methods are indeed able to generalize to experimental data. 
We need to keep in mind, that when moving from the simulated digital twin data to experimental and incorporating the acoustical inversion one will encounter several problems not tackled in this paper. Below, we give a short discussion of these obstacles.

\begin{enumerate}
    \item Classical methods produce inaccurate estimates from the acoustical inversion and learned methods still face issues for out-of-distribution samples. This inaccuracy is directly passed to the optical inversion and to the optical values.

    \item Due to the low amount of experimental samples, it is required to augment the training set with simulated time-series. Simulation of accurate time-series data that correlates well with the experimental data is a challenging topic for ongoing research.

    \item Additional device-specific modeling error that might need compensation.
\end{enumerate}
The focus of this work was to provide a comprehensive methodological overview for solving the optical problem isolated from the acoustic problem. For which we used an accurately simulated digital twin and under the assumption of accurate initial data of the absorbed energy density. Future work will concentrate on combining acoustic and optical inversion for more accurate reconstructions of experimental data.

\section{Conclusions}\label{sec:conclusions}
 In this work, we described, implemented, and examined learned model-based iterative solvers for the nonlinear inverse problem of QPAT using different GD, GN, and SR1-based step directions. Two different training schemes were investigated: (1) the end-to-end scheme, leading to an optimal solution after $K$ trained iterations; and (2) the computationally cheaper greedy scheme, where the iterations were trained separately. The first numerical study revealed the differences between the step directions and training schemes under ideal conditions: In general, the more informative GN step direction performed well with both greedy and end-to-end schemes, whereas the GD and SR1 performed robustly only within the end-to-end training scheme.

In the second study, the digital twin problem was introduced, providing a closer to real-life setup, with a scarce amount of training data and modeling inaccuracies. The results showed that the end-to-end trained GN and GD networks were partly able to compensate for the modeling error.
However, the learned iterative solvers could only reach an accuracy similar to the fully learned method. To justify the further usage of the learned iterative solvers, future work should investigate model error compensation techniques or employ more accurate light transport models.

While we studied the problem of QPAT in this paper, we expect that the presented insights on step directions, training regimes, and model errors are relevant for nonlinear inverse problems in general.

\bibliographystyle{IEEEtran}

\bibliography{source.bib}

\end{document}


\section*{Supplementary Material}

\subsection{Radiative transfer equation for QPAT}
For QPAT, the time-independent RTE in $r\in\Omega \subset \mathbb{R}^n$ $(n=2$ or 3) with given scattering $\mu_s$ and absorption $\mu_a$ coefficients can be written as \cite{ishimaru1978wave}
\begin{equation}\label{eq:RTE}
    \begin{aligned}
    &\hat{s} \cdot \nabla \phi(r, \hat{s})+\left(\mu_s+\mu_a\right) \phi(r, \hat{s})\\
    &=\mu_s \int_{S^{n-1}} \Theta\left(\hat{s} \cdot \hat{s}^{\prime}\right) \phi\left(r, \hat{s}^{\prime}\right) \mathrm{d} \hat{s}^{\prime}+q(r, \hat{s}), \quad r \in \Omega,
    \end{aligned}
\end{equation}
where $\Theta\left(\hat{s} \cdot \hat{s}^{\prime}\right)$ is the scattering phase function and $q(r, \hat{s})$ is the source in $\Omega$, $\partial \Omega$ is the domains boundary, and $\hat{s} \in S^{n-1}$ denotes a unit vector in the direction of interest. With the radiance $\phi(r,\hat{s})$, we can compute the photon fluence as
$\Phi(r)=\int_{S^{n-1}} \phi(r, \hat{s}) \mathrm{d} \hat{s}$

For modeling biological tissues, the scattering phase function is commonly chosen to be the Henyey–Greenstein scattering function \cite{henyey1941diffuse}.
For QPAT, the boundary condition of RTE is often formed by assuming that at the boundary $\partial \Omega$, photons travel in an inward direction only at the source positions.

\subsection{Diffusion approximation and finite element matrices}

By choosing Henyey–Greenstein scattering function \cite{henyey1941diffuse} and assuming a diffusive region $r\in\Omega\subset \mathbb{R}^n$ $(n=2$ or 3) with absorption $\mu_a$ and scattering $\mu_s$ coefficients, the diffusion approximation of RTE \eqref{eq:RTE} is given by (see e.g. \cite{arridgeoptical})
\begin{equation*}
    -\nabla \cdot \kappa(r) \nabla \Phi(r)+\mu_a(r) \Phi(r)=q_0(r), \quad r \in \Omega,
\end{equation*}
where $q_0$ is the light source in $\Omega$, the parameter $\kappa(r) = (n(\mu_a(r) + \mu'_{s}(r)))^{-1}$ is the diffusion coefficient where $\mu'_s =(1-g)\mu_s$ is the reduced
scattering coefficient with anisotropy parameter $-1<g<1$. 
A boundary condition for the DA can be formed by assuming that the total inward-directed photon current is zero at the boundary. Moreover, including the possible refraction mismatch between $\Omega$ and its surrounding medium, the boundary condition for DA can be derived to be \cite{arridgeoptical}
\begin{equation*}
\Phi(r)+\frac{1}{2 \gamma} \kappa(r) A \frac{\partial \Phi(r)}{\partial \hat{n}}= \begin{cases}\frac{I(r)}{\gamma}, & r \in \rho \\ 0, & r \in \partial \Omega \backslash \rho,\end{cases},
\end{equation*}  
where $\hat{n}$ is the outward unit normal vector, $I(r)$ is the boundary photon flux at the source positions $\rho \in \partial\Omega, \gamma$ is a dimension-dependent constant with values 1/$\pi$ in $\mathbb{R}^2$ and 1/4 in $\mathbb{R}^3$. The parameter $A$ accounts for the internal reflection at the boundary $\partial\Omega$. If no boundary refraction occurs, $A$ corresponds to 1. 

Assume that the domain is separated into $P$ non-overlapping elements with $N$ grid coordinates, and that the optical parameters in $\Omega$ are represented as
\begin{equation}\label{eq:basis}
    \begin{aligned}
\mu_a(r) \approx \sum_{t=1}^N \mu_{a_t} \varphi_t(r), \quad
\mu_s(r) \approx \sum_{t=1}^N \mu_{s_t} \varphi_t(r),
\end{aligned}
\end{equation}
with the chosen finite element basis $\varphi_t$. The finite element approximation for the diffusion approximation can then be written as
\begin{equation*}
(M+C+R)\Phi=Q,
\end{equation*}
where 
\begin{equation*}\label{eq:FE_eqs}
    \begin{aligned}
& M(i, j)=\sum_{t=1}^N\frac{1}{n(\mu_{a_t}+\mu^\prime_{s_t})} \int_{\Omega} \varphi_t(r) \nabla \varphi_i(r) \cdot \nabla \varphi_j(r) \mathrm{d} r\\
& C(i, j)=\sum_{t=1}^N\mu_{a_t}\int_{\Omega}  \varphi_t(r)\varphi_i(r) \varphi_j(r) \mathrm{d} r\\
& R(i, j)=\int_{\partial \Omega} \frac{2 \gamma}{A} \varphi_i(r) \varphi_j(r) \mathrm{d} S\\
& Q(i)=\int_{\partial \Omega} \frac{2 I(r)}{A} \varphi_i(r) \mathrm{d} S. 
\end{aligned}
\end{equation*}

\makeatletter 
\renewcommand{\thefigure}{S\@arabic\c@figure}
\makeatother

\subsection{Distribution of simulated optical coefficients}

\begin{figure}[h]
    \centering
    \includegraphics[width=0.95\linewidth]{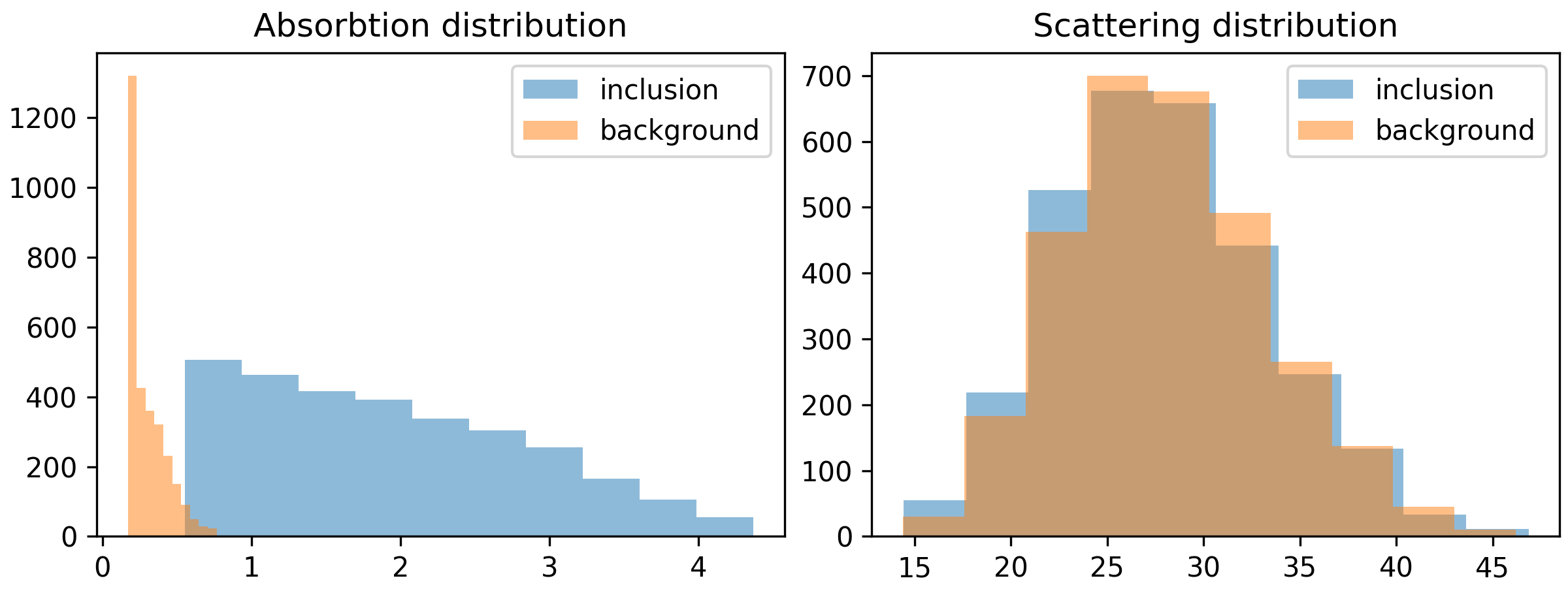}
    \caption{\textit{Digital twin problem}: Distribution of the simulated absorption coefficient (left) and scattering coefficient (right) values for both the background material (orange) and the inclusion materials (blue). While there is a distinct difference between the background and inclusion materials on the absorption values, there is no difference in the scattering distribution. The y-axes shows the bin frequencies when drawing 10,000 samples. The x-axes show the absorption and scattering coefficient in units of cm$^{-1}$.}
    \label{fig:mua_mus_distribution}
\end{figure}

\subsection{Reconstructions from time-series data}
\begin{figure}[h]
    \centering
    \includegraphics[width=1.0\linewidth]{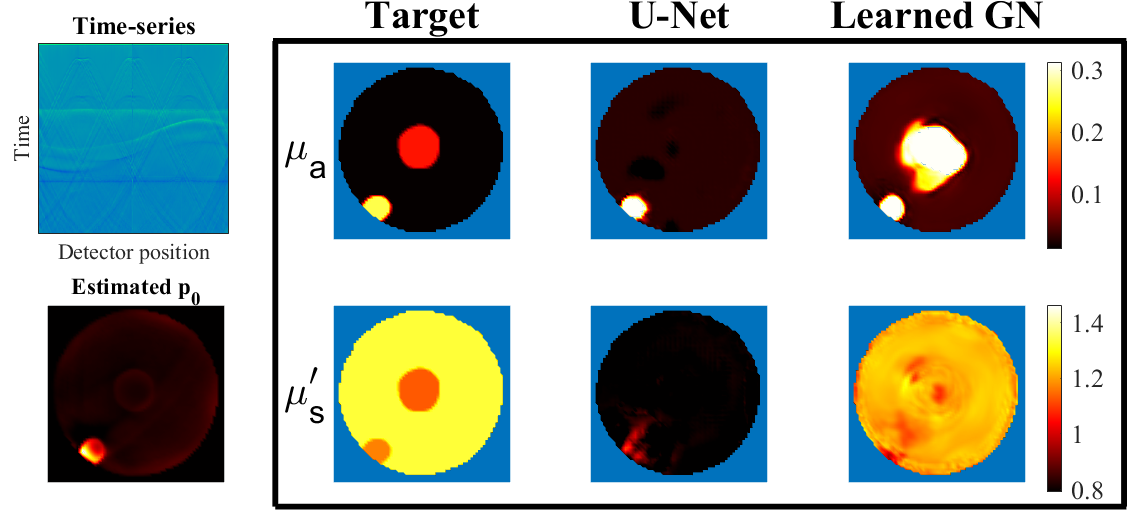}
    \caption{Absorption ($\mu_a$) and reduced scattering ($\mu^\prime_{s}$) reconstructions of a test sample solved from experimental time-series data. The left-most column shows an initial acoustic field $p_0$ (bottom) that is estimated from the shown time-series (top) using a classical iterative solver.} 
    \label{fig:exp_2}
\end{figure}

\begin{figure}[h]
    \centering
    \includegraphics[width=1.0\linewidth]{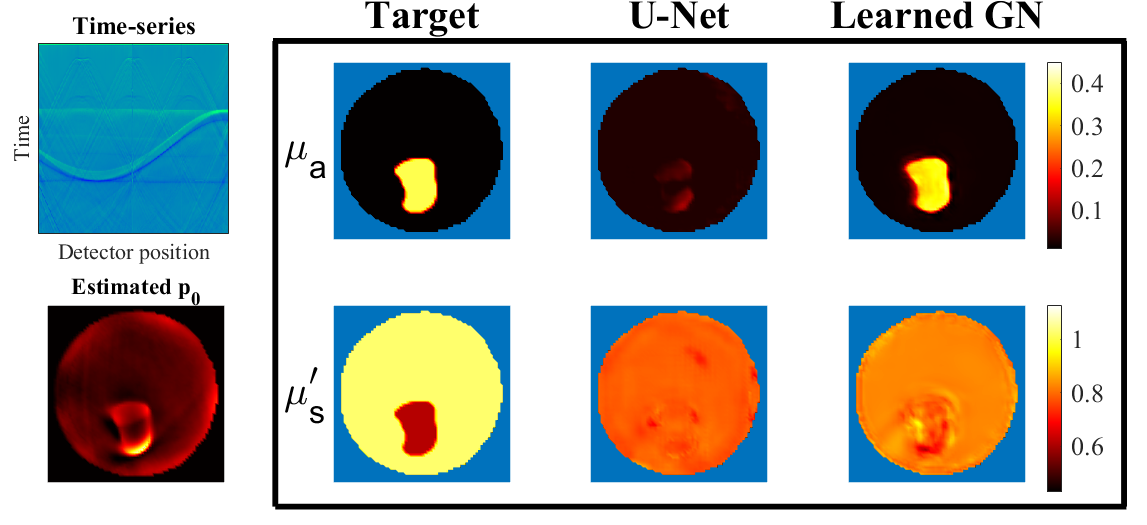}
    \caption{Absorption ($\mu_a$) and reduced scattering ($\mu^\prime_{s}$) reconstructions of a test sample solved from experimental time-series data. The left-most column shows an initial acoustic field $p_0$ (bottom) that is estimated from the shown time-series (top) using a classical iterative solver.} 
    \label{fig:exp_4}
\end{figure}

Here, we present a few additional test reconstructions from the time-series data. Reconstructions of the acoustic problem have been obtained from the corresponding variational problem with TV regularization solved using the primal-dual hybrid gradient, as discussed in \cite{hauptmann2025fast}.
The optical problem was solved using the U-Net and a learned iterative Gauss-Newton solver (Learned GN). Figs.~\ref{fig:exp_2} and \ref{fig:exp_4} show  reconstructions from both solvers.

\bibliographystyle{IEEEtran}

\bibliography{source_sup.bib}